\documentclass[12pt]{article}
\usepackage[dvips]{graphicx}
\begin{document}
\title{On quasinormal modes of small Schwarzschild-Anti-de-Sitter
black hole}
\author{R.A.Konoplya \\
Department of Physics, Dniepropetrovsk National University\\
St. Naukova 13, Dniepropetrovsk  49050, Ukraine\\
konoplya@ff.dsu.dp.ua}
\date{}
\maketitle
\thispagestyle{empty}
\begin{abstract}
We compute the quasinormal modes associated with decay of
the massless scalar filed around a small Schwarzschild-Anti-de-Sitter
black hole. The computations shows that when the horizon radius is much
less than the anti-de-Sitter radius, the imaginary part of the frequency
goes to zero as $r_{+}^{d-2}$ while the real part of $\omega$ decreases
to its minimum and then goes to $d-1$. Thus the quasinormal modes approach
the usual AdS modes in the limit $r_{+} \rightarrow 0$. This
agrees with suggestions of Horowitz et al ({\it Phys.Rev.}
{\bf{D62}} 024027 (2000)).
\end{abstract}

Original interest to quasinormal modes of black holes arose since
they are the characteristics of black holes which do not depend on initial
perturbations and are functions of a black hole parameters only.
At present, this interest is renewed since the QN frequencies are in
the suggested region of the gravitational wave detectors which are
under construction. In general QNM's are important in black holes
dynamics and appear in such processes as collisions of two black
holes, decay of different fields in a BH background.

All this motivated the investigation of the QNM
of black holes in asymptotically flat space-time (see
\cite{Kokkotas-Schmidt} for a recent review).
 The
quasinormal modes of asymptotically de-Sitter black holes were
studied in \cite{Brady-Chambers1},\cite{Brady-Chambers2}. Recently an
unexpected application of quasinormal modes have appeared due to the
AdS/CFT correspondence \cite{Maldacena}: it proved out that a large
black hole in AdS space corresponds to an approximately thermal state
in the CFT, and, thereby, perturbation of the black hole corresponds
to the perturbation of the above thermal state, while the decay of
the perturbation can be associated with the return to thermal
equilibrium. Thus the QN frequencies give us the thermalization
timescale which is very hard to compute directly. Quasinormal modes
for different types of perturbations of black holes in AdS space have
been studied recently in a lot of papers
\cite{Horowitz1}-\cite{Govindarajan-Suneeta}.

Black holes are considered to be small (large), when its horizon radius
is much smaller (larger) then the anti-de-Sitter radius.
When computing the QN frequencies associated with the decay of
massless scalar field in the background of small SAdS BH a striking
conjecture with the black hole critical phenomena was found:
$\omega_{Im}$ is proportional to BH radius $r_{+}$ to high accuracy,
and the slope of the line $\omega_{Im}$ to the $r_{+}$ axis, 2.66,
turned out to be very close
to the special frequency $\lambda=2.67$ which corresponds to the growing
mode $\exp{\lambda t}$ describing the late time behaviour of
the critical solution \cite{Gundlach}.  Yet, the quasinormal frequencies
of the SAdS black
hole have been computed only for black holes with horizon radius
$r_{+} \geq 0.4 R$, where $R$ is the anti-de-Sitter radius
(except for one mode corresponding to $r_{+}=0.2 R$ for which
the behavior of the wave function was numerically reproduced). This is
not sufficient to study the small black hole limit, but several
suggestions were made. In \cite{Govindarajan-Suneeta} it was
stated that both real and
imaginary parts of the QN modes for small black holes are very
large and proportional to the surface gravity, however, later, it was
shown by numerical integration of the wave equation, that at least
for $r_{+} \geq 0.4 R$, in agreement with the first study \cite{Horowitz1},
\cite{Horowitz2}, that both
the real and imaginary parts of $\omega$ are decreasing, and,
stated that the behaviour $ \omega_{Re} \rightarrow const $,
$\omega_{Im} \rightarrow 0$ at $r_{+} \rightarrow 0$ is
expected. Note that by considering an approximate symmetry of the SAdS
metric in the limit  $r_{+} \rightarrow 0$ it was supposed that
$\omega_{Im} \rightarrow 0$ as $r_{+}^2$  \cite{Horowitz1}.
Here we try to compute the quasinormal modes of black holes with
the horizon radius smaller than $0.4 R$,  to
approach the small black holes regime $(r_{+} \ll R)$ as much
as possible, and to obtain more definite hints of
very small black hole behaviour. It proves out that computations
of quasinormal modes for very small black holes are quite
reliable within the method proposed in \cite{Horowitz2} provided one
avoids accumulating of
numerical error (see Appendix).

The d-dimensional Schwarzschild-Anti-de-Sitter metric is:
\begin{equation}\label{100}
ds^{2}= -f(r) dt^{2} + f^{-1}(r)dr^{2} +r^{2}d\Omega^{2}_{d-2},
\end{equation}
where
\begin{equation}\label{101}
f(r)=1-\frac{r_{0}^{d-3}}{r}
+\frac{r^{2}}{R^{2}}.
\end{equation}
Here $r_{0}$ is related to the black hole mass
$$M = \frac{(d-2) A_{d-2} r_{0}^{d-3}}{16 \pi  G_{d}},$$ and $A_{d-2}$ is the
area of a unit (d-2) sphere.

Quasinormal modes of black holes in asymptotically anti-de-Sitter
space time are governed by the wave equation
\begin{equation}\label{1}
\left(\frac{d^{2}}{dr_{\ast}^{2}} + \omega^{2}\right) \Psi(r) = U
\Psi(r),
\end{equation}
where the potential $U$ is given by
\begin{equation}\label{2}
U(r)=f(r)\left( \frac{l(l+d-3)}{r^{2}}-
\frac{(2-d) (d-4)}{4 r^{2}}f(r) +\frac{2-d}{2 r} f'(r)\right),
\end{equation}
and we take $\omega =\omega_{Re}- \imath \omega_{Im}$.
The tortoise coordinate is
$dr^{\ast} = f^{-1}(r)dr$, $l$ is the angular harmonic index. By
rescaling of $r$ we can put $R=1$. It is essential that the effective
potential is infinite at spatial infinity. Thus the wave function
vanishes at infinity and satisfies the purely in-going wave condition
at the black hole horizon.

We managed to computed the quasinormal modes for the $d=4$ black hole with
the radius up to $r_{+}=1/30 R$ (see Fig1-2).
This reasonably approximates
behavior in small black hole regime.  It proved out that the
oscillation frequency falls
down to some minimum, and then begins to grow approaching $d-1$
when $r_{+} \rightarrow 0$ (see Fig 2). This minimum of the
 $Re \omega$ equals
$$min(\omega_{Re})\approx 2.362868 \quad at \quad r_{+}=0.395 R, \quad d=4 $$
$$min(\omega_{Re})\approx 3.705140\quad at \quad r_{+}=0.341 R,  \quad d=5, $$
and (according to our preliminary results) continues to increase
for higher dimensions. Herewith the more dimensional AdS space is the
less black hole radius $r_{+}$ at which the real oscillation frequency
attains its minimum.
Upon thorough consideration of the Fig.1 of the paper \cite{Wang3}
one can learn that among the real oscillation frequencies corresponding
to $r_{+}=0.2, 0.4, 0.6, 0.8R$, $\omega_{Re}$ at $r_{+}=0.4 R$ is the
least. This agrees with our computations showing the minimum
of $\omega_{Re}$ at $r_{+}\approx 0.395 R$ for a four dimensional black hole.
The imaginary part of $\omega$ fall down to zero, and the closer
$r_{+}$ to zero the better the corresponding plot can be fit by
the function $A r_{+}^{2}$. For $d=4$ the best fit for the
last five points (from $r_{+}=1/16$ to $r_{+}=1/30$) in the Tab.1
is $\omega_{Im} = 8.06653 r_{+}^{2}$. The higher the
dimensions are the less $\omega_{Im}$ of small black holes is, i.e.
the more the damping time of a perturbation.

\begin{center}
\begin{tabular}{|c|c|c|}
\hline
  $r_{+}$ & $\omega_{Re}$ & $\omega_{Im}$ \\
\hline
  0.3 & 2.38447 & 0.70413 \\
  0.25 & 2.41945 & 0.54735 \\
  0.2 & 2.47511 & 0.3899 \\
  0.125 & 2.62274 & 0.16392 \\
  0.1 & 2.69282 &  0.10096 \\
  1/12 & 2.74472 & 0.06616 \\
  1/14 & 2.78341 & 0.04578 \\
  1/16 & 2.81289 & 0.03311 \\
  1/18 & 2.83574 & 0.02491 \\
  1/20 & 2.8539  & 0.01932 \\
  1/25 & 2.88584 & 0.01138 \\
  1/30 & 2.9065  & 0.0074 \\ \hline
\end{tabular}
\end{center}
\begin{center}
Tab.1 The fundamental quasi-normal frequencies corresponding
to $d=4$ SAdS black hole, $R=1$.
\end{center}
Thus, even though the boundary conditions at
$r= r_{+}$ do not reduce to regularity at the origin
in the limit $r_{+} \rightarrow 0$, the quasinormal modes
approach the usual AdS modes \cite{Burgess-Lutken} in this limit
(we checked it out for $d=4,5$), i.e.
$\omega_{Re} \rightarrow d-1$ and $\omega_{Im} \rightarrow 0$, as was
discussed in \cite{Horowitz1}.

\begin{figure}
\begin{center}
\includegraphics{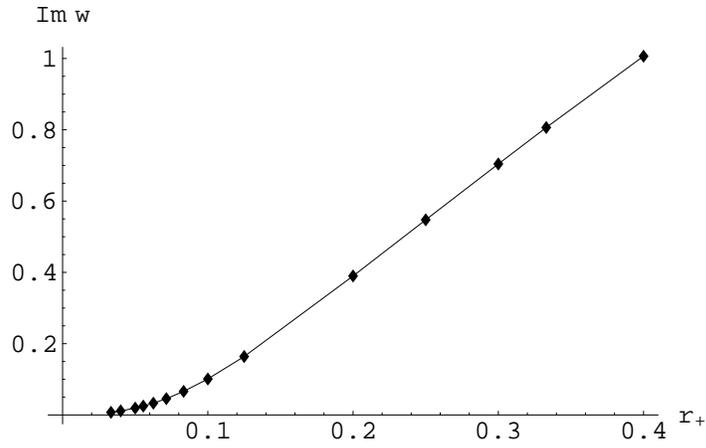}
\caption{Imaginary part of $\omega$ for $d=4$ black hole, $l=0$, $n=0$.}
\label{antides_small1}
\end{center}
\end{figure}

\begin{figure}
\begin{center}
\includegraphics{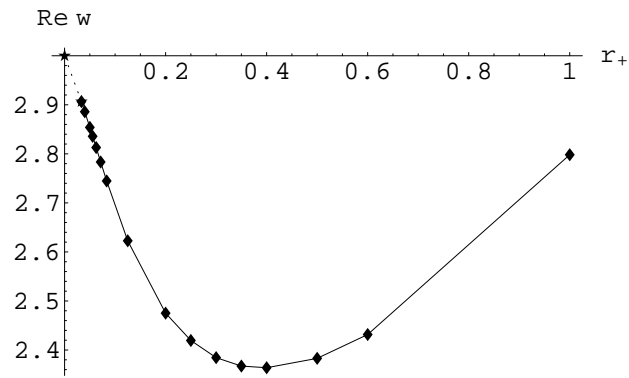}
\caption{Real part of $\omega$ for $d=4$ black hole, $l=0$, $n=0$.}
\label{antides_small2a}
\end{center}
\end{figure}

\begin{figure}
\begin{center}
\includegraphics{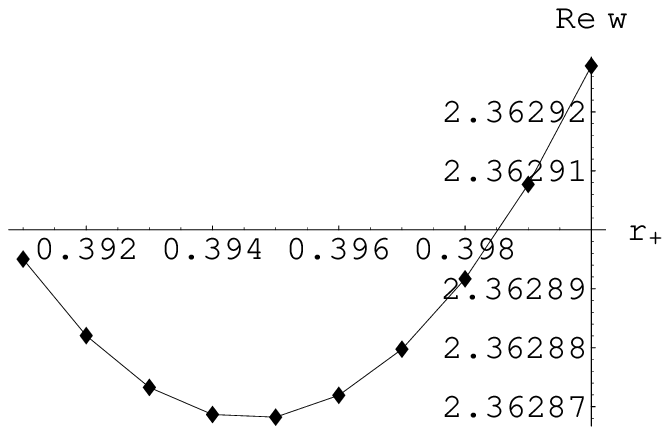}
\caption{Real part of $\omega$ for $d=4$ black hole near its minimum,
$l=0$, $n=0$.}
\label{antides_small7}
\end{center}
\end{figure}

\begin{figure}
\begin{center}
\includegraphics{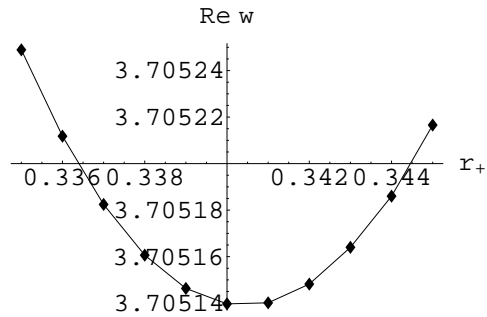}
\caption{Real part of $\omega$ for $d=5$ black hole near its minimum,
$l=0$, $n=0$.}
\label{antides_small8}
\end{center}
\end{figure}

\newpage

\section*{Appendix}

When computing the quasinormal mode one has to truncate the sum representing
the wave function
\begin{equation}\label{3}
\psi(x) = \sum_{n=0}^{\infty}a_{n}(x-x_{+})^{n}, \quad x=\frac{1}{r}
\end{equation}
with some large $N$
and find the roots of the equation $ | \psi(x) |=0$ at $r=\infty (x=0)$.
After simplification the truncated sum (\ref{3}) takes the polynomial
in $\omega$ form, and the necessary roots can easily be found by  {\it
Mathematica}. However this reduction to a polynomial form takes a lot
of computer time and can not be performed for a sum of order $N \sim
200$ or more. Thus for small black holes  we have
to use the trial and error method. Herewith there is a danger of
missing
the fundamental mode we are seeking, and,  of "catching" another overtone.
Fortunately, for reasonably small black holes there are no other overtones
close to the fundamental one, and the minimums of $| \psi |$ are
sufficiently widely separated. Another difficulty
is that for small values of $r_{+}$ the initial tiny errors when
determining the quantities involved in the sum in (\ref{3}) (namely $u_{i}$
, $t_{i}$, and $s_{i}$ of the paper \cite{Horowitz1}) begin to grow when
coming to $N$ of order $1000$ or greater.
Therefore one has to improve precision of these
quantities (with the help of a build-in function of Mathematica) up until
further increasing of precision will not
influence the result. It proved out, that the $50$-digital precision
of $u_{i}$, $t_{i}$, $s_{i}$ is quite enough in this paper.
In addition on must set higher precision of recurrence relations
for coefficients $a_{n}$.
When approaching the limit $r_{+}=0$ the number $N$ of the truncated
sum (\ref{3}) at which an approximate frequency converge grows
as is shown on
the Fig.5 for $d=4$. The more $d$, the more the number $N$ giving  good
approximation of the frequency.
 When following all these receptions
one can be sure that the convergence plot will be smooth and that at
small changing of $r_{+}$ the corresponding frequency will
not change noisily.

\begin{figure}
\begin{center}
\includegraphics{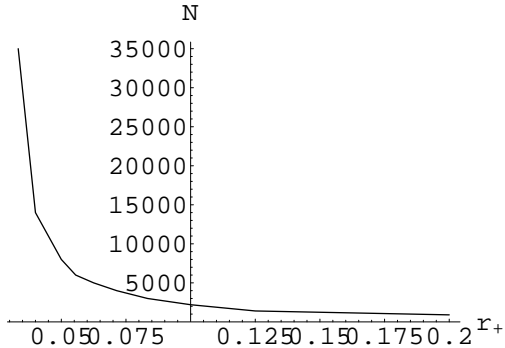}
\caption{Convergence plot for $Re \omega$, $d=4$, $l=0$, $n=0$.}
\label{antides_small6}
\end{center}
\end{figure}

\end{document}